\documentclass[journal = ancac3]{achemso}
\usepackage{graphicx}
\usepackage{color}
\usepackage{amsfonts}

\usepackage{setspace}
\usepackage{xkeyval}

\title{Tunable Carrier Multiplication and Cooling in Graphene} 
\author{Jens Christian Johannsen}
\affiliation{Institute of Condensed Matter Physics, \'Ecole Polytechnique F\'ed\'erale de Lausanne (EPFL), Switzerland}
\author{S\o ren Ulstrup}
\affiliation{Department of Physics and Astronomy, Interdisciplinary Nanoscience Center (iNANO), Aarhus University, Denmark}
\author{Alberto Crepaldi}
\author{Federico Cilento}
\affiliation{Elettra - Sincrotrone Trieste S.C.p.A., 34149 Basovizza, Trieste, Italy}
\author{Michele Zacchigna}
\affiliation{IOM-CNR Laboratorio TASC, Area Science Park,Trieste, Italy}
\author{Jill A. Miwa}
\affiliation{Department of Physics and Astronomy, Interdisciplinary Nanoscience Center (iNANO), Aarhus University, Denmark}
\author{Cephise Cacho}
\author{Richard T. Chapman}
\affiliation{Central Laser Facility, STFC Rutherford Appleton Laboratory, Harwell, United Kingdom}
\author{Emma Springate}
\affiliation{Central Laser Facility, STFC Rutherford Appleton Laboratory, Harwell, United Kingdom}
\author{Felix Fromm}
\affiliation{Lehrstuhl f{\"u}r Technische Physik, Universit{\"a}t Erlangen-N{\"u}rnberg,Germany}
\author{Christian Raidel}
\affiliation{Lehrstuhl f{\"u}r Technische Physik, Universit{\"a}t Erlangen-N{\"u}rnberg,Germany}
\author{Thomas Seyller}
\affiliation{Institut f{\"u}r Physik, Technische Universit{\"a}t Chemnitz , Germany}
\author{Phil D. C. King}
\affiliation{SUPA, School of Physics and Astronomy, University of St. Andrews,
St. Andrews, United Kingdom}
\author{Fulvio Parmigiani}
\affiliation{Elettra - Sincrotrone Trieste S.C.p.A., 34149 Basovizza, Trieste, Italy}
\affiliation{Department of Physics, University of Trieste, Italy}
\author{Marco Grioni}
\affiliation{Institute of Condensed Matter Physics, \'Ecole Polytechnique F\'ed\'erale de Lausanne (EPFL), Switzerland}
\author{Philip Hofmann}
\affiliation{Department of Physics and Astronomy, Interdisciplinary Nanoscience Center (iNANO), Aarhus University, Denmark}
\email{philip@phys.au.dk}

\SectionNumbersOff
\makeatletter
\let\acs@address@list\relax
\makeatother

\begin{document}
\newpage
\begin{abstract}

Time- and angle-resolved photoemission measurements on two doped graphene samples displaying different doping levels reveal remarkable differences in the ultrafast dynamics of the hot carriers in the Dirac cone. In the more strongly ($n$-)doped graphene, we observe larger carrier multiplication factors ($>$ 3) and a significantly faster phonon-mediated cooling of the carriers back to equilibrium compared to in the less ($p$-)doped graphene. These results suggest that a careful tuning of the doping level allows for an effective manipulation of graphene's dynamical response to a photoexcitation.

\end{abstract}

\noindent{}KEYWORDS: Epitaxial graphene, hot carrier dynamics, charge carrier tunability, carrier multiplication, time-resolved ARPES. 

\clearpage

A central question in the field of optoelectronics concerns identifying materials that can convert light into electrical energy with high efficiency, and where an active control of the underlying ultrafast charge carrier dynamics can be achieved. Graphene has emerged as a promising candidate thanks to its unique electronic and optical properties encompassing a very high room-temperature carrier mobility, a truly two-dimensional electronic structure and a constant absorption in the energy range described by a Dirac cone \cite{Geim:2007,Xia:2009a,Bonaccorso:2010}. Recently, a new intriguing effect was added to the list of attractive properties, namely the ability to generate multiple hot carriers with an energy above the Fermi energy from a single absorbed photon as a result of so-called impact excitation processes taking part in the energy relaxation of the primary photoexcited carriers \cite{Winzer:2010,Song:2013,Brida:2013aa,Ploetzing:2014,Tielrooij:2012,Tani:2012}. This carrier multiplication (CM) is predicted to be particularly effective in graphene by virtue of its linear band structure \cite{Rana:2007,Winzer:2010} combined with strong electron-electron (e-e) scattering \cite{kotov:2012,Song:2011d} and weak electron-phonon (e-ph) cooling \cite{Song:2012c,Johannsenb:2013}, and represents a very interesting approach to convert light energy into electronic excitations in an efficient manner. The number of generated hot carriers in the ultrafast cascade of impact excitation processes has furthermore been predicted to be highly sensitive to the doping level of the Dirac carriers, i.e. the location of the Fermi level (FL) relative to the Dirac point (DP) \cite{Song:2013}. Intriguingly, this suggests that this parameter can be used as a knob for an active control of the CM, since the doping level can be easily tuned either electrostatically by gate voltages \cite{Geim:2007,Yu:2009} or chemically by intercalating different elements into the interface in epitaxially grown graphene\cite{Larciprete:2012,Riedl:2009}. The tunability of the ultrafast dynamics of the excited Dirac carriers via the doping level also enables an effective manipulation of graphene's photoconductive properties as a consequence of different carrier scattering rates at different doping levels \cite{Shi:2014, Frenzel:2014}. Moreover, it has been predicted that the cooling rate of the hot carriers back to the equilibrium state displays a strong doping dependence \cite{Bistritzer:2009}.

For realizing practical applications and to motivate further studies, it is of paramount importance to obtain a more comprehensive view of the hot carrier dynamics in the Dirac cone of graphene at different doping levels. Laser-based femtosecond (fs) time- and angle-resolved photoemission spectroscopy (TR-ARPES) is a particularly powerful tool in this respect as it allows the direct observation of the photoexcited electrons' evolution out of thermodynamic equilibrium with energy and momentum resolution. This is a virtue that it shares with no time-resolved optical technique. To date, however, only a few TR-ARPES experiments have been performed on graphene \cite{Johannsenb:2013,Gierz:2013aa,Someya:2014, Ulstrup:2014b}, and none of these addresses the role of doping on the dynamics of the photoexcited electron distribution. In this Letter, we present results from TR-ARPES measurements of photoexcited electrons in graphene under two very different doping conditions. Two epitaxial graphene samples were grown on semiconducting SiC(0001) using two different well-documented synthesis procedures \cite{Riedl:2009,Emtsev:2009} that provide high-quality monolayer graphene displaying either $n$- or $p$-type doping. The doping mechanisms are different as the interface layer decoupling the graphene samples from the SiC substrate is different in the two samples \cite{Ristein:2012};  a carbon-rich layer for $n$-doped and a hydrogen layer for $p$-doped graphene. In the $n$-doped graphene sample, the DP is 380 meV below the FL corresponding to a carrier concentration of 1 $\times 10^{13}$ cm$^{-2}$ whereas in the $p$-doped case a carrier concentration of 4 $\times 10^{12}$ cm$^{-2}$ places the DP 240 meV above the Fermi energy. 

The $n$-doped sample is thus more doped than the $p$-doped sample. Our direct mapping in momentum space of the transient occupation of the Dirac cone clearly shows that this difference in doping level, renders the hot carrier dynamics in the two graphene samples remarkably different to the extent that a substantial CM is possible in the more strongly ($n$-)doped case.

Before proceeding, a note on terminology is appropriate. The electron-hole symmetry of the low-energy electronic structure of graphene implies an equivalence in the dynamics of electrons and holes. As we with TR-ARPES, however, only reliably can measure electrons, no distinction, unless expressly stated, between \emph{carriers} and \emph{electrons} is made in this manuscript and we use these terms interchangeably.
 
\begin{figure}[h!tbp]
\includegraphics[width=.99\textwidth]{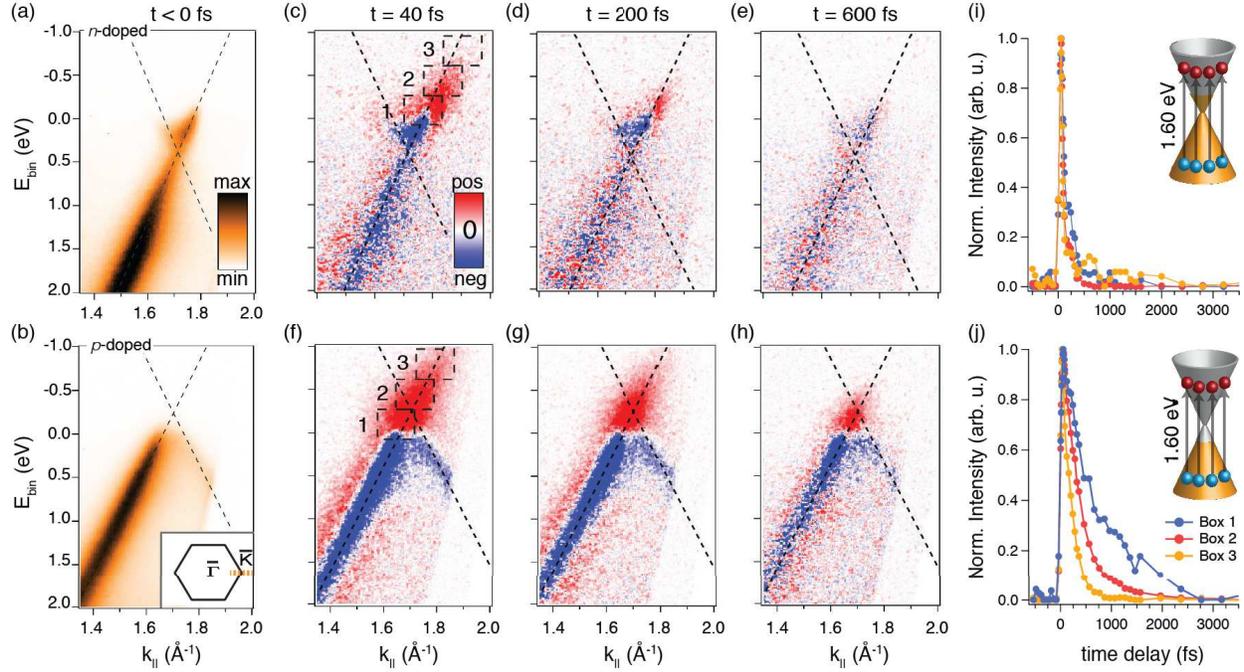}\\
\caption{Comparing the relaxation dynamics in $n$- and $p$-doped graphene samples: (a, b) Photoemission intensity around the $\bar{K}$-point ($\bar{\Gamma}-\bar{K}$ direction) in (a) the $n$- and (b) the $p$-doped graphene before the arrival of the pump pulse. The dispersion obtained from a tight binding calculation has been added as a guide to the eye (dashed line). (c,d,e) and (f,g,h) ARPES difference spectra for the $n$- and the $p$-doped sample, respectively, obtained by subtracting the spectra in (a) and (b) from spectra acquired at the indicated time delays. (i, j) Normalized spectral intensity binned in the three boxed regions in (c) for (i) the $n$-doped and in (f) for (j) the $p$-doped graphene. The boxed regions are placed at the same binding energies in the two cases. Schematics of the excitation with the applied pump energy of 1.6 eV are shown in the insets. Electrons and holes are depicted as red and blue spheres, respectively.}
\label{fig:1}
\end{figure}

The TR-ARPES experiments were performed at the ARPES end-station at the Artemis facility, Rutherford Appleton Laboratory \cite{Cacho:2014}. The high-energy, ultrafast pulses provided by high-harmonic generation in an argon gas enable acquisition of TR-ARPES data from the high-momentum states at the $\bar{K}$-point of the Brillouin zone of graphene ($|\bar{\Gamma}\bar{K}|$ = 1.7 \AA ). After \emph{in-situ} cleaning of the samples in ultrahigh vacuum by annealing to 600 K, TR-ARPES data were acquired by the spatially overlapping pump ($h \nu_{P}$ =1.60 eV) and probe \mbox{($h \nu_{EUV} = 21$ eV)} beams. The pump fluence was set to 1.4 mJ/cm$^2$ focused to a 400 $\mu$m diameter spot (full width at half maximum) on the sample, and both the probe and the pump beam were polarized perpendicular to the scattering plane that was oriented along the $\bar{\Gamma} -\bar{K} $ direction. We note here that this experimental geometry implies that we are probing along a maximum of the initially anisotropic photo-injected carrier distribution \cite{Malic:2011} that re-distributes itself into an isotopic distribution on the time scale of the carrier thermalization.\cite{Malic:2011,Brida:2013aa,Mittendorf:2014}. The sample temperature was fixed at 300 K, and the total time, energy and angular resolution were better than 40 fs, 200 meV and 0.3$^{\circ}$, respectively.     

Fig.~\ref{fig:1}(a) and (b) show ARPES data from the states in the Dirac cone of the $n$- and the $p$-doped graphene sample, respectively, acquired before pump excitation, i.e. at negative time delays. For the $n$-doped graphene, the DP is discernible via a decrease in photoemission intensity. In the $p$-doped graphene, the DP is situated above the FL as indicated by the dashed linear band that is the result of a fit of a tight-binding model to the data. The effect of bringing the electronic system out of its thermodynamic equilibrium state can be visualized by taking the difference between each ARPES spectrum acquired for positive time delays and the equilibrium spectrum taken at a negative time delay. In Fig.~\ref{fig:1}(c)-(e) and Fig.~\ref{fig:1}(f)-(h), we show the results of this subtraction for the $n$- and the $p$-doped graphene, respectively, for three selected positive time delays (40 fs, 200 fs and 600 fs). In both the $n$- and $p$-doped case, a strong increase (decrease) in spectral intensity is clearly visible at 40 fs in the conduction (valence) band. This redistribution of spectral weight reflects the optical excitation of the Dirac carriers, and it is noticeable that it is rather uniform along the band and not centered around $h \nu_{P}/ 2$ (800 meV) above the DP, i.e. at the binding energy of the final states in the optically excited inter-band transition. This observation fits well with the consensus in the literature that a significant part of the relaxation dynamics occurs already during the action of the ultrashort pump pulse in the form of efficient e-e scattering events filling the states between the thermal and the initial non-thermal carrier distribution \cite{Winzer:2010,Malic:2011,Song:2013,Brida:2013aa}. The excited carrier density is observed to be larger in the $p$-doped graphene than in the $n$-doped graphene. This may appear puzzling because the optical transition takes place between the same filled and empty states in both cases (see insets in Fig.~\ref{fig:1}(i) and Fig.~\ref{fig:1}(j)) and one should thus expect a similar density of excited carriers. Note, however, that the possibility for an excitation as such is restricted by the presence of already excited electrons in the final state, an effect also observed in degenerated differential transmission studies \cite{Sun:2010c,George:2008}. This so-called Pauli blocking implies that light absorption is particularly efficient if the excited state can already be emptied during the pump pulse \cite{Winzer:2012a}. Given the short time scale for Auger transitions, this is also possible, but it is much more effective for the $p$-doped sample with more available phase space for the decay of the electrons below the final state of the photoexcitation. 

Following the subsequent temporal evolution of the excited carrier density at larger time delays (Fig.~\ref{fig:1}(d)-(e) and Fig.~\ref{fig:1}(g)-(h)), it appears that the decay back to the equilibrium state is significantly faster in the $n$-doped than in the $p$-doped sample. At 600 fs, the spectrum from the $n$-doped sample in Fig.~\ref{fig:1}(e) already appears close to fully relaxed while an excess population of carriers is still present about the DP in the $p$-doped sample, as seen in Fig.~\ref{fig:1}(h). This is corroborated by tracking the intensity as a function of time delay in three distinct boxes in the conduction band, as shown in Fig.~\ref{fig:1}(c) and (f). The intensities as a function of time delay are shown in Fig.~\ref{fig:1}(i)-(j), and indicate that the relaxation dynamics taking place after the pump excitation indeed is much faster in the $n$-doped than in the $p$-doped sample.

\begin{figure}
\includegraphics[width=1.0\textwidth]{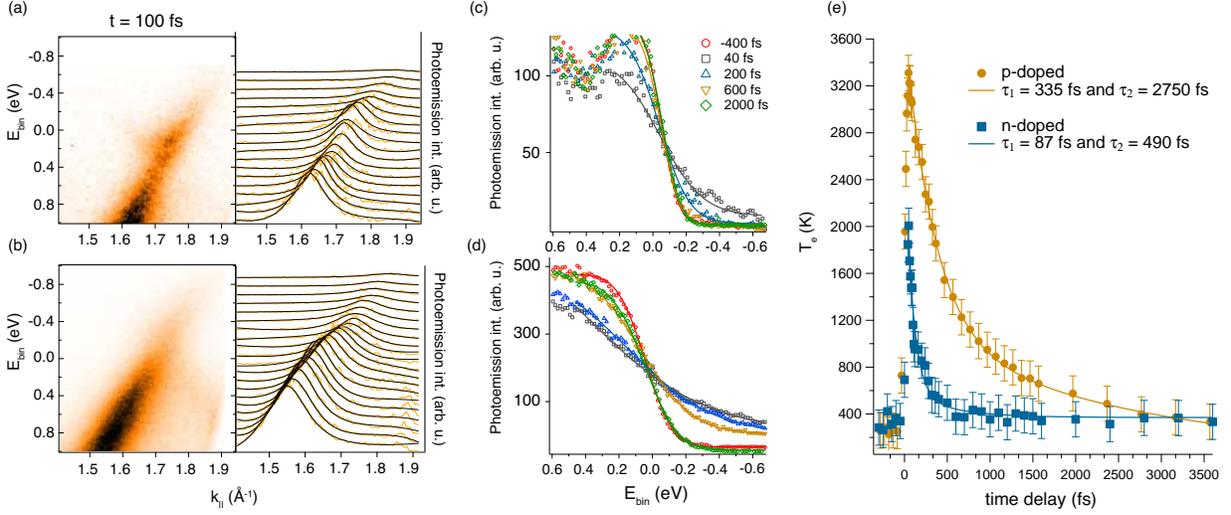}
\caption{Extracting the electronic temperature in the $n$- and $p$-doped graphene sample. (a,b) Left panel: ARPES spectrum measured at 100~fs after optical excitation in (a) $n$- and (b) $p$-doped graphene. Right panel: The same spectrum decomposed into momentum distribution curves (MDCs) for which only a subset is shown. The results of the fits by Lorentzian line shapes to each individual MDC are represented by the black solid lines. (c,d) Integral of fitted MDC peak intensities in (c) the $n$- and (d) the $p$-doped graphene at selected time delays fitted by Fermi-Dirac distributions. (e) The fitted electronic temperature (with error-bars added) for the $n$- (blue markers) and the $p$-doped graphene (amber markers) as a function of time delay. Lines are the result of double exponential function fits to the data. The obtained time constants are indicated.}
\label{fig:2}
\end{figure}

Great care should be exercised when performing a comparison of the spectrally resolved dynamics in this manner due to the non-linearity of the Fermi-Dirac (FD) distribution \cite{Crepaldi:2012}. A conceptually more appealing approach would be to compare the evolution of the electronic temperature characterizing the hot carrier population. This is possible provided that a thermal FD distribution is established within the pump pulse duration. The experimental access with energy and momentum resolution to the transient occupation of states provided by our TR-ARPES data allows us to directly extract the time-dependent temperature of the hot electrons in distinct states using a simple method described in details elsewhere \cite{Johannsenb:2013,ulstrupc:2014}. Briefly, we decompose the ARPES spectrum at all time delays into momentum distribution curves (MDCs), each representing the photoemission intensity at a given binding energy as a function of $k_{\|}$.  An example of such a decomposition is shown in Fig.~\ref{fig:2}(a) and (b) for the $n$- and the $p$-doped graphene, respectively. All MDCs are then fitted to Lorentzian functions and a polynomial background, and the intensity under the fitted line shape is momentum-integrated and finally plotted against the binding energy. This gives access to the photoemission intensity of the band independent of its dispersion as displayed in Fig.~\ref{fig:2}(c) and (d). The data show that the thermalization of the electronic systems occurs on a few tens of fs in both the $n$- and the $p$-doped case, although a small deviation from a FD distribution is present in the former at a time delay of 40 fs, see Fig.~\ref{fig:2}(c). This is in agreement with our previous study on $p$-doped graphene that also shows an ultrafast redistribution of carriers towards a hot FD distribution \cite{Johannsenb:2013}. 

The maximum temperature reached after the thermalization (i.e for t $> 40$ fs) is significantly lower in the $n$-doped (2000 K) than in the $p$-doped graphene (3300 K), as seen in Fig.~\ref{fig:2}(e). We note here that a laser pump fluence of 1.4 mJ/cm$^2$ and an optical absorption coefficient for graphene on SiC of 1.3 $\%$ \cite{Drabi:2010} give a nominal absorbed fluence of 18.2 $\mu$J/cm$^2$. This amount of absorbed energy, however, would imply substantially higher temperatures than observed for either samples due to the small electronic specific heat of graphene (see the Supporting Information). A similar discrepancy between applied pump fluence and actual observed electronic temperature has been reported in a THz spectroscopy study on graphite \cite{Kampfrath:2005} and is also observed in another recent TR-ARPES study \cite{Gierz:2013aa}. Several reasons can be envisaged to account for the observed lower values. We believe that the most important effects are Pauli blocking during the pump excitation and a non-optimal geometric overlap between the optical spots and the focus of the electron analyzer \cite{George:2008,Sun:2010c}. A recent theoretical study reports a saturation fluence due to Pauli blocking of 0.65 mJ/cm$^2$ for pump pulses with a duration of a few tens of femtoseconds \cite{Winzer:2012a}. With a pump fluence of 1.4 mJ/cm$^2$, we are thus in the fluence regime where Pauli blocking is expected to be efficient. In fact, similar dynamics and absolut values for the electronic temperature in both samples are observed when we pump with higher fluences, as shown in the Supporting Information. Given that we can directly extract the maximum temperature reached by the carriers from the TR-ARPES data and know the electronic specific heat, we can estimate that the effective fluence in the $n$- and $p$-doped graphene samples are 0.4 $\mu$J/cm$^2$ and 1.5 $\mu$J/cm$^2$, respectively (see the Supporting Information for more details on this calculation). 

For both the $n$- and the $p$-doped graphene, we observe a bi-exponential temperature decay back to the equilibrium state. The first decay is governed by energy dissipation via the emission of optical phonons. This cooling channel experiences a bottleneck when the hot carriers' energy drops well below the unusually large Debye energy ($\approx$ 200 meV), at which point the second decay sets in characterized by the emission of acoustic phonons assisted by disorder-scattering (so-called supercollision processes) \cite{Song:2012c, Johannsenb:2013,Betz:2012,Graham:2012}. Both decay time constants are substantially faster in the $n$-doped case, as indicated in Fig.~\ref{fig:2}(e). An explanation for this difference in carrier cooling efficiency between the two samples is related to the fact that the cooling dynamics should quantitatively scale with the available phase space for the hot carriers to scatter into after their interaction with the phonon modes \cite{Hwang:2008}. The different doping levels exhibited by the two samples render this phase space different in the two cases. In the $p$-doped graphene, where the FL is 240 meV below the DP, the tail of the hot FD distribution crosses the region about the DP where the density of states vanishes. As a result, all cooling processes involving intra-band optical and, in particular, acoustic phonon scattering events (including supercollisions) towards this region are significantly slowed down. A bottleneck for the carrier cooling is thereby created in a similar manner as in a real semiconductor with a sizable band gap. This is not the case in the $n$-doped graphene as the doping level for this sample is larger and, as a result, the hot holes and electrons do not cascade through the DP in their relaxation process. Hence, the phase space for decay of the hot carriers in the $n$-doped graphene is not constrained to the same extent as in the $p$-doped case, and the probability for finding an unoccupied state at lower energy after having emitted a phonon is significantly larger; in many ways reminiscent of the situation in a real metal. It is important to reiterate that due to the electron-hole symmetry in the linear part of the electronic structure of graphene, there exist no fundamental difference between $n$- and $p$-doped graphene; the difference in the carrier cooling efficiency observed here is entirely due to the different degree of doping or, equivalently, the density of carriers. This density dependence of the cooling rates is in qualitative agreement with the theoretical study by R. Bistritzer and A.H. MacDonald \cite{Bistritzer:2009}, but does not agree on a quantitative level, as the latter study does not take supercollisions into account, resulting in cooling rates on the order of nanoseconds. 

\begin{figure}
\includegraphics[width=0.5\textwidth]{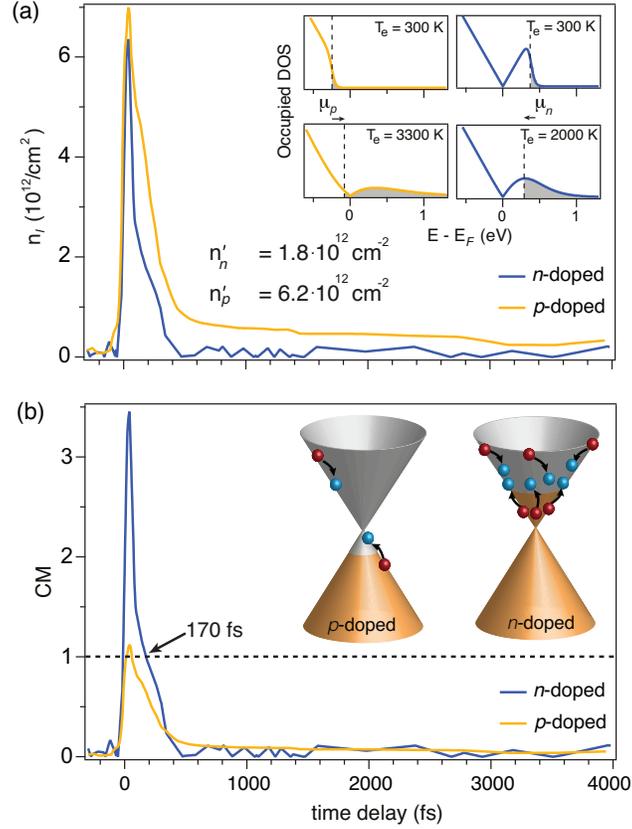}\\
\caption{Quantifying the carrier multiplication in the $n$- and the $p$-doped graphene. (a) Time-resolved density of optically induced excited carriers $n_I$ calculated from the extracted electronic temperature. The inset demonstrates how $n_I$ is found by integration (gray-shaded area under the curves) of the occupied density of states (DOS), i.e. the DOS multiplied by the FD distribution, above the temperature- and doping-dependent chemical potential $\mu_{p(n)}$. Note that in order to conserve overall charge neutrality, the chemical potential is moving towards the Dirac point for higher temperatures in both $n$- and $p$-doped graphene. The optically injected density of carriers $n_{p(n)}^{\prime}$ for the $p$($n$)-doped graphene is provided. (b) Carrier multiplication (CM) from Eqn.~\ref{eq:cm} in the main text. The impact excitation processes leading to CM are illustrated by arrows for the $n$- and the $p$-doped Dirac cones in the inset, where electrons are depicted as red spheres and holes as blue spheres. }
\label{fig:3}
\end{figure}

The direct access to the absolute value and dynamics of the electronic temperature $T_e (t)$ enables us to quantify the CM factor in the $n$- and the $p$-doped graphene. We note that throughout this work the term CM is used for all inter - and intra-band scattering processes that increase the number of hot electrons in the conduction band \cite{Tielrooij:2012,Brida:2013aa}, despite the fact that inter-band (intra-band) processes are dominant in $p$($n$) doped graphene. We determine the number of photoexcited carriers generated by the pump pulse $n_{p(n)}^{\prime}$ as the ratio between the total absorbed energy needed to heat the carriers from 300 K to their maximum temperature and the pump photon energy. From $T_e (t)$, we calculate the time-dependent total density of hot carriers promoted to states above the FL, $n_I (T_{e}(t))$. In this calculation, the temperature-dependent position of the FL (chemical potential) is explicitly taken into account as explained in more detail in the Supporting Information. The result of these calculations can be seen in Fig.~\ref{fig:3}(a). We observe that the total hot carrier density above the FL $n_I (T_e(t))$ is almost the same in the two samples for short time delays, despite the carrier density injected by the pulse $n_{p(n)}^{\prime}$ being a factor of 3 less in the $n$-doped sample. This is a consequence of the larger density of states around the FL in $n$-doped compared to the $p$-doped graphene, as shown in the insets and discussed above. The CM factor is given by the following ratio \cite{Winzer:2010}:
\begin{eqnarray}
\mathrm{CM}=\frac{n_I (T_e(t))-n_I (300K)}{n^{\prime}}
\label{eq:cm}
\end{eqnarray}
where we have subtracted the thermal carrier background $n_I$ (300K) from the total carrier density. The determined CM for the $n$- and the $p$-doped graphene is plotted in Fig.~\ref{fig:3}(b) as a function of time delay. Intriguingly, the CM is more than a factor of 3 larger in the $n$-doped case. The Dirac cone diagrams in the inset provide a simple explanation for this result: in the $n$-doped graphene the photoexcited carriers can easily find scattering partners in the Fermi sea to involve in impact excitation processes as both the number of nearby thermal carriers and the scattering phase space is large. These processes are on the other hand severely constrained in the $p$-doped graphene due to the vanishing density of states at the DP. The lower electronic temperature in the $n$-doped graphene will also favor a higher CM as the larger occupation gradient around the FL lowers the rate of the inverse Auger recombination processes thereby creating an imbalance between impact excitation and Auger recombination \cite{Winzer:2012}.

The large value for the CM of $\approx3.5$ found at the end of the initial almost instantaneous thermalization process is a very intriguing result indeed, as this indicates that the energy of the photoexcited carriers is efficiently harvested during this process by the electronic system. This result is consistent with the joint experiment-theory study by Pl\"otzing \emph{et al.} \cite{Ploetzing:2014} that reports CM factors larger than two for our effective fluence range. Subsequently, phonon-emitting recombination processes become the dominant relaxation channel for the additional generated hot carriers that are shown to persist above the FL for approximately 170~fs. This sets an upper limit for the time available to extract the carriers in an optoelectronic device desiring to exploit CM. In the $p$-doped sample, the CM is shown to reach values above unity only for short time delays. This is apparently in contrast with our previous TR-ARPES study of this effect on a $p$-doped graphene sample displaying the same level of doping \cite{Johannsenb:2013}, where a CM $> 1$ was not observed. We note, however, that a higher time resolution is achieved in the present experiment. Moreover, the excitation energy of the pump pulse is larger which is expected to increase the CM as long as the energy remains within the linear regime of the band structure \cite{Winzer:2012}.
  
With TR-ARPES, we can directly follow the ultrafast dynamics in the electronic band structure of a material. In this work, we take advantage of this ability to show in a direct manner that the dynamical response of the Dirac carriers in graphene to a photoexcitation can be tuned by varying the doping level, i.e. the position of the FL relative to the DP. In particular, a substantial CM reaching values larger than three can be achieved in $n$-doped graphene doped to a level of 380 meV. This is significantly larger than the CM reachable in $p$-doped graphene displaying a lower level of doping of 240 meV. The subsequent cooling of the hot carriers via phonon-emitting processes is  significantly more efficient in the $n$-doped graphene compared to in the $p$-doped graphene as a consequence of the reduced phase-space for decay imposed by the vanishing density of states at the DP. Taken together, these results suggest that a careful tuning of the doping level in graphene in between the two levels displayed by our samples allows for a significant generation of multiple hot carriers. For photovoltaic applications, it is interesting to note that the solar radiation reaching the Earth on a femtosecond time scale is orders of magnitude lower ( $\sim$ pJ/cm$^2$) than in our experiment, and thus in a low fluence regime, where CM is predicted to be even more efficient as observed here. Finally, we note that the tunability of the carrier dynamics by doping and fluence is also important for reaching efficient carrier heating regimes\cite{Jensen:2014}.

\begin{acknowledgement}

We gratefully acknowledge financial support from the VILLUM foundation, The Danish Council for Independent Research, the Lundbeck Foundation, the Swiss National Science Foundation (NSF), EPSRC, The Royal Society and the Italian Ministry of University and Research (Grants No. FIRBRBAP045JF2 and No. FIRB-RBAP06AWK3). Access to the Artemis Facility was funded by STFC. Work in Erlangen and Chemnitz was supported by the European Union through the project ConceptGraphene, and by the German Research Foundation in the framework of the SPP 1459 Graphene.

\end{acknowledgement}

\providecommand{\latin}[1]{#1}
\providecommand*\mcitethebibliography{\thebibliography}
\csname @ifundefined\endcsname{endmcitethebibliography}
  {\let\endmcitethebibliography\endthebibliography}{}

\newpage

\end{document}